\documentclass{jfm}

\usepackage{amsmath,amsfonts,amssymb,stmaryrd,amsbsy,bm}
\usepackage{upmath}
\usepackage{mathtools}
\usepackage{graphicx}
\usepackage{hyperref}
\usepackage{breakurl}
\usepackage{accents}
\usepackage{natbib}
\usepackage[normalem]{ulem}

\makeatletter
\def\squiggly{\bgroup \markoverwith{\textcolor{red}{\lower3.5\p@\hbox{\sixly \char58}}}\ULon}
\makeatother

\newcommand\redout{\bgroup\markoverwith{\textcolor{red}{\rule[.5ex]{2pt}{1.4pt}}}\ULon}

\renewcommand{\thesection}{\arabic{section}}

\newcommand{\traction}{\bm{f}}
\newcommand{\force}{\bm{F}}
\newcommand{\torque}{\bm{L}}

\newcommand{\uu}{{ \bm u}}
\newcommand{\vv}{{ \bm v}}
\newcommand{\xx}{{ \bm x}}
\newcommand{\rr}{{ \bm r}}
\newcommand{\UU}{{ \bm U}}

\newcommand{\Tnsr}[1]{\bm{\mathsfbi #1}} 

\newcommand{\Kt}{\Tnsr{E}}
\newcommand{\Kr}{\Tnsr{G}}
\newcommand{\kt}{\mathcal E}
\newcommand{\kr}{\mathcal G}

\newcommand{\Rstar}{R^*}
\newcommand{\calO}{\mathcal{O}}

\newcommand{\surface}{S}

\newcommand{\evec}[1]{{\bm{e}}_{#1}}
\newcommand{\nvec}{{\bm{n}}}

\newcommand{\proj}[1]{{\,\bm{\Tnsr{P}}}_{\!\!#1}}

\newcommand{\fnal}[1]{[#1]}

\newcommand{\vslip}{{\bm v}_{\text{sl}}}
\newcommand{\Urigid}{\bm{\mathcal U}_{\bm{U\Omega}}}

\begin{document}
\title{Bypassing slip velocity: rotational and translational velocities of autophoretic
colloids in terms of surface flux}
\author{Paul E. Lammert\aff{1}
\corresp{\email{lammert@psu.edu}},
Vincent H. Crespi\aff{1,2,3},
Amir Nourhani\aff{1}
\corresp{\email{nourhani@psu.edu}}
}

\affiliation{
\aff{1}Department of Physics, Pennsylvania State University, University Park, PA 16802
\aff{2}Department of Materials Science and Engineering, The Pennsylvania State University, University Park, PA 16802
\aff{3}Department of Chemistry, The Pennsylvania State University, University Park, PA 16802
}

\maketitle
\date{June 16, 2016}

\begin{abstract}
A standard approach to propulsion velocities of autophoretic colloids with
thin interaction layers uses a reciprocity relation applied to the slip velocity.
But the surface flux (chemical, electrical, thermal, etc.), which is the source 
of the field driving the slip is often more accessible. We show how, under 
conditions of low Reynolds number and a field obeying the Laplace equation in 
the outer region, the slip velocity can be bypassed in velocity calculations. 
In a sense, the actual slip velocity and a normal field proportional to the flux 
density are equivalent for this type of calculation.
Using known results for surface traction induced by rotating or translating an
inert particle in a quiescent fluid, we derive simple and explicit integral formulas
for translational and rotational velocities of arbitrary spheroidal and slender-body 
autophoretic colloids.
\end{abstract}

\section{Introduction\label{sec:intro}}

In recent years, several varieties of autophoretic colloidal particles 
have been fabricated and studied in the laboratory~\citep{Paxton+-04,Gibbs+Zhao-09,Jiang+-10,Ebbens+Howse-10,Wang+-13-Rev}.
Under common approximations \citep{Anderson89} including thinness of 
the interfacial layer near the particle surface $\surface$, the small Reynolds number 
self-propulsion of such a particle is understood in terms of a slip velocity 
$\vslip = \mu \bnabla_{\!\surface\,} \Phi$ generated across the interfacial layer
by the tangential gradient $\bnabla_{\!\surface\,}$ of a field $\Phi$ --- electric 
potential (electrophoresis), chemical concentration (diffusiophoresis, electrophoresis), 
or temperature (thermophoresis) --- obeying the Laplace equation in the outer region
when P\'eclet number is small.
Although the slip mobility $\mu$ can vary with position, we take it uniform,
as is commonly done.
From $\vslip$, the particle velocity can be found via a (Lorentz) reciprocity relation,
if the surface traction generated by translating an inert particle in quiescent fluid
is known.
Compared to the classical subject \citep{Anderson89} of phoresis of 
passive particles driven by an externally imposed field $\Phi$, 
the distinctive feature of autophoresis is that $\Phi$ is ultimately due to a flux density 
$J$ at the particle surface of chemical species, heat, etc., which is proportional
to the normal derivative of $\Phi$ and often more accessible than $\bnabla_\surface\Phi$ 
both experimentally and theoretically.
Thus, formulas relating particle velocity and angular velocity directly to 
the flux are highly desirable. Previous formulas of this sort
\citep{Popescu+-10,Yariv-11,Nourhani+-15a,Schnitzer+Yariv-15,Golestanian+07}
have been limited to bodies of axisymmetric shape with an axisymmetric
flux distribution. (Equivalently, only the component of velocity 
along the symmetry axis was found.)
Except for the work \citep{Yariv-11,Schnitzer+Yariv-15} on slender bodies, 
these results have mostly taken the form of expansions in special functions,
which are not always transparent, and can make the identification of asymptotic
limits difficult and tricky, as in~\citep{Popescu+-10}.
We prove [Eq. (\ref{eq:equivalence})] that, within the simple autophoretic model
described above, for arbitrary particle shape, the hydrodynamic force and torque 
generated by the slip velocity is exactly the same as would be generated by a 
hypothetical radial velocity proportional to the flux density!
Thus, the latter can be substituted for the former in the reciprocity method velocity 
formulas, obviating the need to calculate $\Phi$.
Using this result, we easily derive simple integral kernels transforming arbitrary 
flux distributions into the complete rotational and translational velocities of 
both spheroids and slender bodies, recovering the results of \citep{Nourhani+Lammert-16a}
for the former and \citep{Schnitzer+Yariv-15} for the latter.
Simple integral kernels such as those derived here are very valuable for
completely mapping out motor performance over well-defined design spaces.

The body of the paper is structured as follows.
In Section \ref{sec:general-theory}, we present the general theory, 
reviewing (Section \ref{sec:Lorentz}) the use of Lorentz reciprocity for Stokes flow, 
and demonstrating (Section \ref{sec:shortcut}) the central claim that the 
hydrodynamic force and torque generated 
by $\vslip$ are exactly the same as are generated by $(\mu/{\mathcal D})J\nvec$,
with $\mathcal D$ a transport coefficient appearing in the Neumann
boundary condition $J = -{\mathcal D} \partial\Phi/\partial n$.
In Section \ref{sec:axisymmetric}, this result is applied to shape-axisymmetric 
bodies, for which a simple formulation in terms of one-dimensional integrals
is worked out using symmetry. 
Symmetry considerations also show that an autophoretic particle cannot rotate
about its symmetry axis, absent symmetry breaking by the environment, or possibly
an inhomogeneous slip mobility $\mu$.
Methods based on reciprocity require the surface traction on 
a rigidly moving inert particle as input. Using literature results for that,
the scheme is applied to spheroids, both prolate and oblate 
(Section \ref{sec:spheroid}), as well as slender bodies (Section \ref{sec:slender-body})
to derive, in just a few lines, complete and simple integral expressions for
the translational (\ref{eq:spheroid-V}, \ref{eq:slender-body-V}) and rotational
(\ref{eq:spheroid-Omega}, \ref{eq:slender-body-Omega}) velocities.
The reader interested only in the results can skip straight to those, after
a glance at Section \ref{sec:model} and the preamble to Section \ref{sec:axisymmetric}, 
as well as (\ref{eq:Gamma-Fourier}, \ref{eq:Jperp}).
In the concluding section, we observe that the velocity formula for
a slender body suggests that non-convex shapes can propel in a direction
counter to naive expectations.

\section{General theory}\label{sec:general-theory}

This section commences with a more precise definition of our model, followed 
by a review of the use of the Lorentz reciprocity theorem for Stokes flow,
then the main result embodied in Eq. (\ref{eq:equivalence}), which rests on
the key observation (\ref{eq:key-identity}).

\subsection{Model}
\label{sec:model}

Our model consists of a boundary value problem for a fluid-filled, 
unbounded domain $\calO$ with boundary $\partial\calO = S$. 
The surface $S$ --- meant to represent the ``outer edge'' of the infinitely thin 
interfacial layer around an autophoretic particle --- is taken to be a smooth 
closed compact two-manifold embedded in ${\mathbb R}^3$. 
The particle is the source of a field $\Phi$, obeying $\nabla^2\Phi=0$ in $\calO$, and 
with boundary conditions
\begin{subequations}
\label{eq:Phi-BCs}
\begin{align}
& \frac{\partial\Phi}{\partial n}\Big|_S = -\frac{J}{{\mathcal D}},
\\
& \Phi \to \mathrm{constant} \;\text{as}\;\, |x| \to \infty.
\end{align}
\end{subequations}
These reflect the idea that the particle is the only source or sink of $\Phi$.
The flux density $J$ is taken as given in this model, rather than determined
from more basic data, such as chemical kinetics\citep{Sabass+Seifert-12,Nourhani+-15c-PRE}.

Since we are interested in a low-Reynolds number flow, the fluid in $\calO$ is
taken to be governed by the Stokes system 
\begin{equation}
  \label{eq:Stokes}
\eta \nabla^2 {\vv} = \nabla p; \quad \mathop{\mathrm{div}} {\vv} = 0. 
\end{equation}
The boundary conditions on the fluid velocity are 
\begin{subequations}
\begin{align}
&\vv|_S = \vslip = \mu \bnabla_{\!\surface\,} \Phi, 
\label{eq:v-BC-S}  \\
&\vv \to 0 \;\text{as}\;\, |x|\to\infty.
\label{eq:v-BC-infinity}
\end{align}
\end{subequations}
Some auxiliary Stokes flows considered in the following discussion will not 
obey the boundary condition (\ref{eq:v-BC-S}), but they will all obey 
(\ref{eq:v-BC-infinity}).
It is well known 
\citep[Arts. 335--336]{Lamb},\citep[\S 3-2]{Brenner-64a,Happel+Brenner},\citep[\S 4.2]{Kim+Karrila}
that this boundary condition, with compact $S$, implies that the velocity is 
$O(1/|x|)$ and the stress, $O(1/|x|^2)$ as $|x|\to\infty$. 

\subsection{Lorentz reciprocity for Stokes flows}
\label{sec:Lorentz}

The stress tensor for a Stokes field ($\vv$, $p$ pair) is given by
(superscript `$\dagger$' denotes transpose) 
\begin{equation}
\Tnsr{T} = - p\, \Tnsr{I} + \eta \left[ \bnabla\vv + (\bnabla\vv)^\dagger \right].
\end{equation}
An arbitrary pair of Stokes flows $\vv$ and $\uu$ in a bounded volume ${\mathcal V}$
with smooth boundary $\partial{\mathcal V}$ satisfies the well-known reciprocity relation 
\citep{Brenner-64d,Kim+Karrila,Pozrikidis,Happel+Brenner} (`Lorentz reciprocal theorem') 
\begin{equation}
\int_{\partial {\mathcal V}} \vv\cdot  \traction[\uu]\, d\surface  
=   \int_{\partial {\mathcal V}} \uu \cdot \traction[\vv]\, d\surface,
\nonumber
\end{equation}
where $\traction[\vv] := \nvec\cdot \Tnsr{T}|_{\partial {\mathcal V}}$ is the 
hydrodyamic surface force density arising from the flow $\vv$ 
(acting from the side pointed to by $\nvec$). 
In the context of our problem, the reciprocity relation can be applied to the
part of $\calO$ inside a sphere of large volume $R$. But, because of the falloff
implied by boundary condition (\ref{eq:v-BC-infinity}), the integral over that 
sphere vanishes as $R\to\infty$, leaving simply
\begin{equation}
\int_{S} \vv\cdot  \traction[\uu]\, d\surface  
=   \int_{S} \uu \cdot \traction[\vv]\, d\surface.
\label{eq:LRR}
\end{equation}
Now, with $\nvec$ pointing into $\calO$, the net hydrodynamic force $\force$ and torque 
$\torque$ acting across $\surface$ from the outside by the flow $\uu$ are given by
\begin{equation}
\force[\uu] = \int_\surface \traction[\uu]\, d\surface; \qquad
\torque[\uu] = \int_\surface {\bm r}\times \traction[\uu] \, d\surface.
\label{eq:force+torque}
\end{equation}
In the special case that $\uu = \Urigid$ reduces to a rigid-body motion 
\begin{equation}
\Urigid|_S = \UU + {\bm\Omega}\times{\bm r}
\label{eq:U-rigid}
\end{equation}
on $S$, the corresponding surface force density must, by linearity, take the form
\begin{equation}
\label{eq:K-matrices}
 \traction[\Urigid](\xx)
= \Kt(\xx) \cdot \UU + \Kr(\xx) \cdot {\bm \Omega},\ \ \ \
({\bm x}\in\surface),
\end{equation}
for tensor functions $\Kt(\xx)$ and $\Kr(\xx)$. 
Inserting these expressions into the reciprocity relation (\ref{eq:LRR}),
and pulling the arbitrary constants $\UU$ and ${\bm\Omega}$ out of the integrals
yields
\begin{equation}
\label{eq:Force-Torque-K}
\force[\vv] = \int_\surface \Kt^\dagger \cdot \vv\, d\surface;
\qquad
\torque[\vv] = \int_\surface \Kr^\dagger \cdot \vv\, d\surface.
\end{equation}
In particular, if $\int_S f[\Urigid]\cdot (\vv -\vv^\prime)\, dS = 0$ 
for every $\bm{U}$ and $\bm{\Omega}$, then
$\force[{\vv}] = \force[\vv^\prime]$ and $\torque[{\vv}] = \torque[\vv^\prime]$.

Returning to the problem of the motion of an autophoretic particle, we decompose
the fluid velocity at the outer edge of the interfacial layer
into the slip velocity $\vslip$ and an unknown rigid-body motion:
\begin{equation}
\label{eq:particle-V+slip}
\vv
= 
\vslip 
+ \UU_\text{p} + ({\bm \Omega}_\text{p} \times {\bm r})
\quad \text{on }\surface.
\end{equation}
Assuming we know $\Tnsr{E}$ and $\Tnsr{G}$, (\ref{eq:Force-Torque-K}) can be used 
to determine $\UU_{\text{p}}$ and ${\bm\Omega}_{\text{p}}$.
They are whatever is required to provide a force and torque cancelling 
$\force[\vslip]$ and $\torque[\vslip]$, namely,
\begin{equation}
\label{eq:V-Omega-from-v-slip-abstract}
\begin{pmatrix}
\UU_{\text{p}} \\
{\bm\Omega}_{\text{p}}
\end{pmatrix} 
= 
-\begin{pmatrix}
\Tnsr{A} & \Tnsr{B} \\
\Tnsr{B}^\dagger & \Tnsr{C}
\end{pmatrix}^{-1}   
\begin{pmatrix}
\force[\vslip] \\
\torque[\vslip]
\end{pmatrix}. 
\end{equation}
The block matrix here is the symmetric hydrodynamic resistance matrix~\citep{Kim+Karrila},
with blocks given by
\begin{equation}
\label{eq:resistance-matrix-blocks}
\Tnsr{A} = \int_\surface \Kt^\dagger \, d\surface; \quad
\Tnsr{B} = \int_\surface -\Kt^\dagger\times{\bm r} \, d\surface; \quad
\Tnsr{C} = \int_\surface -\Kr^\dagger\times{\bm r} \, d\surface.
\end{equation}

\subsection{A shortcut}
\label{sec:shortcut}

Now, our slip velocity is $\vslip = \mu \bnabla_S\Phi$. If we had $\Phi$ in hand,
(\ref{eq:force+torque},\ref{eq:V-Omega-from-v-slip-abstract}, 
and \ref{eq:resistance-matrix-blocks}) could be used to find the translational
and rotational velocities of the autophoretic particle.
However, the source flux density $J = -{\mathcal D} \partial\Phi/\partial n|_S$
is usually much more accessible, so we would like an expression directly
in terms of $J$, thus avoiding the need to solve for $\Phi$.
The key to this is the identity
\begin{equation}
\label{eq:key-identity}
\int_\surface
\traction[\Urigid]
\cdot \bnabla \Phi\, d\surface =0,
\end{equation}
where $\Urigid$ goes to zero at infinity (\ref{eq:v-BC-infinity}) and 
reduces to $\UU + \bm{\omega} \cdot {\bm r}$
on $S$ (${\omega}_{ik} = \sum_j \epsilon_{ijk}\Omega_j$),
while $\Phi$ obeys the Laplace equation and the boundary conditions (\ref{eq:Phi-BCs}).
To see this, note first that $\bnabla^2\Phi = 0$ guarantees that
$(\vv = \bnabla\Phi, p = 0)$ is a legitimate Stokes flow with stress tensor 
$\Tnsr{T} = 2\eta  \bnabla \bnabla\Phi$;
As $r\to\infty$, $\vv = O(1/r^2)$ and $\Tnsr{T} = O(1/r^3)$.
The reciprocity relation (\ref{eq:LRR}) is therefore applicable, and yields
\begin{equation}
 \int_\surface 
\traction[\Urigid]
\cdot \bnabla\Phi\, d\surface 
 = 
 \int_\surface  \traction[\bnabla\Phi] \cdot 
\Urigid
 \, d\surface
 = 
 \int_\surface  \nvec \cdot\Tnsr{T}\cdot \Urigid \, d\surface.
\label{eq:step1}
\end{equation}
Insert the explicit form of $\Urigid$ on $S$ to rewrite this as
\begin{equation}
\cdots =  \UU\cdot \int_S \nvec \cdot \Tnsr{T} \, dS
+ \int_S \nvec \cdot \Tnsr{T} \cdot \bm{\omega}\cdot\bm{r} \, dS.
\label{eq:step2}
\end{equation}
Now, apply the divergence theorem to obtain
\begin{equation}
\cdots = 
\UU\cdot \int_\calO  
\bnabla \cdot \Tnsr{T} \, dV
+
\int_\calO \bnabla \cdot (\Tnsr{T} \cdot \bm{\omega}\cdot \rr)  \, dV.
\label{eq:step3}
\end{equation}
This step is a bit delicate. Since the integral of 
$\nvec \cdot \Tnsr{T}$ over a sphere of large radius $R$ is
$O(R^2\cdot R^{-3}) = O(R^{-1})$, the conversion of the first integral is legitimate.
For the second one, note that
\begin{equation}
\int_{S_R} \nvec \cdot (\bnabla\bnabla \Phi) \cdot \bm{\omega}\cdot\bm{r} \, dS
= \omega_{jk}\int_{S_R} R n^i n^k \partial_i\partial_j \Phi\, dS.
\nonumber
\end{equation}
The monopole term $1/r$ does not contribute because $\omega_{jk}$ is 
antisymmetric. (Ultimately, this vanishing comes down to the monopole field and 
the rigid rotation field belonging to different representations of
$SO(3)$.)
Moving to the $O(1/r^2)$ dipole contribution shows the
integral to be $O(R\cdot R^{-2}\cdot R^{-2}\cdot R^2) = O(R^{-1})$.
Thus, (\ref{eq:step3}) is justified, and the first integral there is even zero, 
because $\bnabla \cdot \Tnsr{T} =0$. 
Finally, since $\bm{\omega}$ is constant, while $\Tnsr{T}$ 
is divergence-free,
\begin{equation}
\bnabla \cdot (\Tnsr{T} \cdot \bm{\omega}\cdot \rr) =  
\mathrm{Tr}\, (\Tnsr{T}\cdot{\bm \omega}). 
\nonumber
\end{equation}
But, $\Tnsr{T}$ is symmetric, $\bm{\omega}$ anti-symmetric, so this is zero, and
the second integral in (\ref{eq:step3}) with it.
Eq. (\ref{eq:key-identity}) is therefore proved.

The velocity field $\bnabla\Phi$ in the preceding is a purely 
auxiliary entity, introduced for the purpose of obtaining (\ref{eq:key-identity}), 
which can now be used to obtain the result we really need.
Since $\bnabla \Phi|_S = \mu^{-1} \vslip - \nvec {\mathcal D}^{-1} J$, 
the comment immediately following (\ref{eq:Force-Torque-K}) implies that
\begin{align}
\force [\vslip]
&=\frac{\mu}{{\mathcal D}} \force [J \nvec]
\label{eq:equivalence}
\\
\torque [\vslip]
&=\frac{\mu}{{\mathcal D}} \torque [J \nvec].
\nonumber
\end{align}
We could hardly be more fortunate. We wished to work with $J$ instead of $\vslip$,
and these equations grant permission to do so in nearly the most straightforward sense
imaginable: simply replace $\vslip$ in 
(\ref{eq:V-Omega-from-v-slip-abstract}) with $(\mu/{\mathcal D})J\nvec$.
That gives us
\begin{equation}
\label{eq:final-abstract-V-Omega}
\begin{pmatrix}
\UU_{\text{p}} \\
{\bm\Omega}_{\text{p}}
\end{pmatrix} 
= 
-\frac{\mu}{{\mathcal D}}
\begin{pmatrix}
\Tnsr{A} & \Tnsr{B} \\
\Tnsr{B}^\dagger & \Tnsr{C}
\end{pmatrix}^{-1}   
\int \begin{pmatrix} \nvec\cdot\Kt \\ \nvec\cdot\Kr \end{pmatrix} J \, dS.
\end{equation}
Perhaps the most important advantage is that both ${\bm U}_{\text{p}}$ 
and ${\bm \Omega}_{\text{p}}$ are accessible for arbitrary $J$, not just the 
axisymmetric flux distributions heretofore treated.
To use this reciprocity-based method, whether directly with $J$, or
with $\vslip$, requires knowledge of the tensor functions $\Tnsr{E}$ and $\Tnsr{G}$
that come from solution of an auxiliary problem involving an inert particle 
rotated and translated in an otherwise quiescent fluid.
The next section takes up that issue.

\section{Axisymmetric bodies\label{sec:axisymmetric}}

Now we apply the general theory of the previous section to {\em shape}-axisymmetric 
bodies (no symmetry assumed of the flux density).
The surface $\surface$ of such a body is given in cylindrical coordinates 
$(z,\rho,\phi)$ by 
\begin{equation}
  \label{eq:axisymmetric-surface}
\surface:\quad
-1\le z\le 1;
\quad 
0\le \phi  < 2\pi;
\quad
\rho=R(z).
\end{equation}
By choice of units, the length of the body is 2, leaving the radius 
function $R: [-1,1] \rightarrow (0,\infty)$ as the only variable element
(undercuts are not allowed).
In many cases, as for the spheroids and slender bodies treated below,
one wants a family of surfaces obtained by varying
a scaling paramter $\kappa$:
\begin{equation}
R(z) = \kappa \Rstar(z).
\end{equation}

In Section \ref{sec:symmetry}, we develop some general formulae for 
the rotational and translational velocities of axisymmetric bodies. 
They are applied in Sections \ref{sec:spheroid} and \ref{sec:slender-body}
to the spheroid family and slender bodies, respectively, using literature
results for the surface traction on an inert translating and rotating particle.

\subsection{Symmetry and  reduction to one dimension}
\label{sec:symmetry}

Now we use $C_\infty$ rotational symmetry about the $z$-axis and the
attendant mirror symmetries to simplify the general problem of determining
translational and rotational velocities. 
For axisymmetric bodies generally, the translational and rotational
problems can be decoupled, all the required integrals reduce to one-dimensional 
integrals over $z$ involving a handful of functions characterizing
the hydrodynamic properties of $\surface$ and only three Fourier components
of $J$ (with respect to $\phi$).

Decoupling of the translational and rotational problems is accomplished by
finding a point about which pure rotations entail no net force. With respect
to that {\it center of resistance}, the off-diagonal blocks 
$\Tnsr{B}$, $\Tnsr{B}^\dagger$ of the
resistance matrix (\ref{eq:resistance-matrix-blocks}) vanish.
Recall that, under reflection in a plane, the perpendicular components of 
ordinary vectors, notably velocity and force, change sign while in-plane 
components are unchanged. On the other hand, components of pseudovectors such as
angular velocity and torque behave in the opposite way.
Consider rotation about the $z$-axis.
By rotational symmetry, the resulting force must be along $z$.
Consideration of a mirror plane containing the $z$-axis shows that it is actually zero.
Now consider rotation $\bm{\Omega}$ about a point $p$ on the $z$-axis.
Consideration of a plane containing $\evec{z}$ and $\bm{\Omega}$ shows that
$\bm{F} \propto \evec{z}\times{\bm{\Omega}}$, with a proportionality 
that changes sign as $p$ moves from $z \ll 0$ to $ 0 \ll z$. By continuity,
there is an intermediate point where it vanishes, which is the sought-for
center of resistance.
In case the body has a reflection plane perpendicular to the
$z$ axis, as for a spheroid, the center of resistance is necessarily 
in that plane.
We show later that for a slender body, the center of resistance is
asymptotically at the midpoint of the body's length.
From now on, we implicitly work with the origin at the center of resistance.
The block $\Tnsr{A}$ is independent of origin, and therefore can be calculated without
knowing where it is.

The tensor functions $\Kt$ and $\Kr$ can be expanded on the
dyadic products $\proj{ij} := \evec{i}\evec{j}$ made from 
$\evec{z}$, $\evec{\rho}$ and $\evec{\phi}$, with coefficients which are 
functions solely of $z$. But, reflection symmetry about planes containing
the $z$-axis forces some coefficients to be zero. Since $\Kt$ transforms vectors
to vectors, it cannot couple components in the $\evec{z}\wedge\evec{\rho}$ plane
to those perpendicular to it, namely $\evec{\phi}$. 
$\Kr$, on the other hand, couples (angular velocity) pseudo-vectors to 
(force) vectors.
Thus, we can write the expansions 
\begin{align}
\Kt\! &= \!
\kt_{zz} \proj{z} 
\!+\! \kt_{\rho\rho} \proj{\rho}
\!+\! \kt_{\phi\phi} \proj{\phi}  
\!+\!\kt_{z\rho} \proj{z\rho} 
\!+\!\kt_{\rho z}  \proj{\rho z},
\nonumber \\
\Kr &= 
 \kr_{z\phi}\proj{z\phi}
+ \kr_{\phi z}\proj{\phi z}
+ \kr_{\rho\phi}\proj{\rho\phi}
+ \kr_{\phi \rho}\proj{\phi \rho}, 
\label{eq:K-expansions}
\end{align}
where $\proj{z}$ abbreviates the orthogonal projector $\proj{zz}$, and
similarly for $\proj{\rho}$ and $\proj{\phi}$.

Having eliminated $\Tnsr{B}$ by choice of origin, symmetry implies that
the remaining blocks of the resistance matrix take the forms
\begin{align}
\label{eq:A+C-form}
{\Tnsr{A}} = {\mathcal A}_z \proj{z} &+ {\mathcal A}_\perp \proj{\perp},
\nonumber \\
{\Tnsr{C}} = {\mathcal C}_z \proj{z} &+ {\mathcal C}_\perp \proj{\perp}.
\end{align}
Substituting (\ref{eq:K-expansions}) into 
Eqs.~(\ref{eq:resistance-matrix-blocks}) and using the fact that the
angular averages of $\evec{\rho}$ and $\evec{\phi}$ are zero, while those of 
$2\proj{\rho}$ and $2\proj{\phi}$ are $\proj{\perp}:=\Tnsr{I}-\proj{z}$ yields
expressions
\begin{subequations}
\label{eq:A+C-components}
\begin{align}
{\mathcal A}_z &= \int \kt_{zz}\, 2\pi R d\ell,\quad
{\mathcal A}_\perp = 
\frac{1}{2} \int (\kt_{\rho\rho} + \kt_{\phi\phi}) \, 2\pi R d\ell,
\label{eq:A-components} \\
{\mathcal C}_z &= \int  R \kr_{\phi z} \, 2\pi R d\ell,
\,\,
{\mathcal C}_\perp = \frac{1}{2} \int  
[ z (\kr_{\rho\phi} - \kr_{\phi\rho}) - R\kr_{z \phi} ] \, 2\pi R d\ell.
\label{eq:C-components}
\end{align}
\end{subequations}
Here, we have written the surface area element as
\begin{equation}
dS = d\phi\, Rd\ell,
\end{equation}
where $d\ell = \sqrt{1+(R^\prime)^2} dz$ is differential arc length along a 
constant-$\phi$ longitudinal section.

To use Eq. (\ref{eq:final-abstract-V-Omega}) for the rotational and translational
velocities, we need to put
\hbox{$\int \nvec\cdot{\Tnsr E}\, J\, dS$} and 
\hbox{$\int \nvec\cdot{\Tnsr G}\, J\, dS$} in the same format.
To do that, Fourier expand the flux density $J(z,\phi)$ with respect to $\phi$,
obtaining
\begin{equation}
  \label{eq:Gamma-Fourier}
J(z,\phi) = J_0(z) + J_x(z)\cos\phi + J_y(z)\sin\phi + \cdots. 
\end{equation}
Only the explicit terms here are needed, more conveniently,
$J_0$ and the vector defined by
\begin{equation}
{\bm J}_\perp(z) := J_x(z) \evec{x}  + J_y(z) \evec{y}.
\label{eq:Jperp}
\end{equation}
This is because those are all that occur in the angular averages at
fixed $z$,
$\langle J \rangle_\phi = J_0(z)$,
$\langle \evec{\rho} J\rangle_\phi = {\bm J}_\perp(z)/2$,
$\langle \evec{\phi} J \rangle_\phi = \evec{z} \times{\bm J}_\perp(z)/2$, 
Unlike in (\ref{eq:A+C-components}) we will take $z$ as integration variable
rather than $\ell$, using
\begin{equation}
\frac{d\ell}{dz}\nvec =  -R^\prime \evec{z} + \evec{\rho}.
\end{equation}
With the expansions (\ref{eq:K-expansions}), we find
\begin{align}
\frac{d\ell}{dz} \nvec\cdot \Tnsr{E} &= 
(-R^\prime \kt_{zz} + \kt_{\rho z})\, \evec{z} 
+ (-R^\prime \kt_{z\rho} + + \kt_{\rho\rho})\, \evec{\rho},
\nonumber \\
\frac{d\ell}{dz} \nvec\cdot \Tnsr{G} &= 
(-R^\prime \kr_{z\phi} + \kr_{\rho\phi})  \evec{\phi}.
\nonumber
\end{align}
Insertion into Eqs.~(\ref{eq:Force-Torque-K}) produces
\begin{align}
\label{eq:F-from-k}
\frac{{\mathcal D}}{\mu}
{\force\fnal{\nvec J}}
&  =
{2\pi}
\int \Big[ 
( \kt_{\rho z} -R^\prime \kt_{zz} )
J_0 \, \evec{z} 
+ (\kt_{\rho\rho} - R^\prime \kt_{z\rho} )\frac{{\bm J}_\perp}{2} \Big] 
R\, dz 
\nonumber \\
\frac{{\mathcal D}}{\mu}
{\torque\fnal{\nvec J}}
 & = 
{2\pi}{\evec{z}} \times
\int 
\frac{{\bm J}_\perp}{2}
(-R^\prime \kr_{z\phi} + \kr_{\rho\phi})  
\, R\, dz.
\end{align}
Combining this with (\ref{eq:A+C-components}), we finally obtain
\begin{subequations}
\label{eq:U+Omega-final}
\begin{align}
\frac{{\mathcal D}}{\mu}
\UU_{\text{p}} &= 
\frac{2\pi}{{\mathcal A}_z}
{\evec{z}}
\int ( R^\prime \kt_{zz} - \kt_{\rho z} ) J_0 \, R dz
+ \frac{\pi}{{\mathcal A}_\perp}  
\int {{\bm J}_\perp} 
(R^\prime \kt_{z\rho}- \kt_{\rho\rho} ) R\, dz,
\label{eq:U-final} \\
\frac{{\mathcal D}}{\mu}
{\bm \Omega}_{\text{p}} &=
\frac{\pi}{{\mathcal C}_\perp} 
{\evec{z}} \times
\int 
{{\bm J}_\perp}
(R^\prime \kr_{z\phi} - \kr_{\rho\phi})  
\, R\, dz.\hfill
\label{eq:Omega-final}
\end{align}
\end{subequations}

An interesting consequence of (\ref{eq:Omega-final}) is that the particle cannot generate 
an angular velocity about its shape symmetry axis $\evec{z}$. Actually, this is implied by
symmetry and linearity, and is therefore independent of the thin boundary layer approximation.
By linearity, if it were possible for the particle to rotate about $\evec{z}$, 
some single Fourier component of $J$ would suffice, say $J \propto \cos(m\phi)$.
But, $\evec{x}\wedge\evec{z}$ is a mirror plane for the surface decorated with the 
scalar field $J$ or the vector field $\nvec J$, while the proposed
pseudovector $\Omega$ lies within this plane. 
An autophoretic sphere, therefore, ought not to rotate at all, regardless of $J$.
If it does, it must be due to a symmetry-breaking environment or nonuniform
slippability $\mu$.

\subsection{The spheroid family}
\label{sec:spheroid}

The spheroidal family of surfaces is generated by the standard radius
\begin{equation}
\label{eq:spheroid-family}
\Rstar(z) = \sqrt{1-z^2},\quad\quad 0 < \kappa.
\end{equation}
If $\kappa < 1$ ($\kappa=1$, $\kappa > 1$), 
this describes a prolate spheroid with eccentricity $\varepsilon = \sqrt{1-\kappa^2}$ 
(sphere, oblate spheroid with eccentricity $\sqrt{1-\kappa^{-2}}$).
For a spheroid \citep{Brenner-64d,Fair+Anderson},
\begin{equation}
  \label{eq:spheroid-Kt}
\Kt = (\nvec\cdot{\bm r}) \, (\alpha \proj{z} + \beta \proj{\perp}),
\end{equation}
with ${\bm r}$ denoting position relative to the center of the body,
and $\alpha$, $\beta$ constants (values of which will not be needed).
In terms of components (\ref{eq:K-expansions}),
\begin{equation}
\kt_{zz} = \alpha (\nvec\cdot{\bm r}),\,\,  
\kt_{\rho\rho} = \kt_{\phi\phi}  = \beta (\nvec\cdot{\bm r}).
\end{equation}

From (\ref{eq:spheroid-family}), simple manipulations lead to
\begin{align}
& \frac{d\Rstar}{dz} = -\frac{z}{\Rstar}, 
\nonumber \\
& \frac{d\ell}{dz} = \frac{1}{\Rstar} \sqrt{1+(\kappa^2-1)z^2},
\nonumber \\
& \nvec = \frac{\kappa z\evec{z} + \sqrt{1-z^2}\, \evec{\rho}}{\sqrt{1+(\kappa^2-1)z^2}},
\nonumber \\
& (\nvec\cdot{\bm r})  R \, d\ell = \kappa^2 dz.
\end{align}
Insertion into (\ref{eq:U-final}) gives the particle velocity
  \begin{align}
\label{eq:spheroid-V}
\UU_{\text{p}} 
& = 
\frac{\mu}{{\mathcal D}}
\int
\left(
\kappa \frac{d\Rstar}{dz}J_0(z) \evec{z} 
- \frac{1}{2}{\bm J}_\perp(z)  
\right) \frac{dz}{d\ell}
\frac{dz}{2}
\nonumber \\
& = 
-\frac{\mu}{{\mathcal D}}
\int
\left(
\kappa {z}J_0(z) \evec{z} 
+ \frac{\Rstar}{2}{\bm J}_\perp(z)  
\right) 
\frac{dz}{ 2\sqrt{1+(\kappa^2-1)z^2} }
  \end{align}
The second form here is more practical; the first faciliates comparison with the 
slender body result (\ref{eq:slender-body-V}).
The case of fully axisymmetric $J$ has been 
studied \citep{Popescu+-10,Nourhani+-15b}, but the complete formula
(\ref{eq:spheroid-V}) does not seem to be in the literature.

In a similarly automatic way, the angular velocity is computed using \citep{Fair+Anderson}
\begin{equation}
\Kr = (\nvec\cdot{\bm r}) \, \left\{
-\alpha R \proj{\phi z}
 +\beta [z(\proj{\phi\rho} - \proj{\rho\phi}) + R \proj{z\phi}] 
\right\},
\nonumber
\end{equation}
with $\alpha$, $\beta$ (possibly new) constants. Only the three coefficient functions
\begin{equation}
\kr_{\phi\rho} = -\kr_{\rho\phi} 
= \beta z (\nvec\cdot{\bm r}),\quad
\kr_{z \phi} = \beta R (\nvec\cdot{\bm r})
\label{eq:contradiction}
\end{equation}
are actually needed.
Plugging into (\ref{eq:Omega-final}) yields
\begin{equation}
\label{eq:spheroid-Omega}
{\bm \Omega}_{\text{p}} = -\frac{3}{4}
\frac{\mu}{{\mathcal D}}
\left(\frac{1-\kappa^2}{1+\kappa^2}\right)
\evec{z} \times \int
{\bm J}_\perp(z) \frac{dz}{d\ell} \, z dz.
\end{equation}
Note that, in accordance with earlier discussion, this vanishes for $\kappa = 1$ (sphere).

\subsection{Slender bodies}
\label{sec:slender-body}

In this section we develop an asymptotic theory which imposes no particular form 
for $\Rstar$, but applies, {\it a priori}, only in the limit of small $\kappa$.
A radius function of a slender body family is
\begin{equation}
  \label{eq:slender-body}
\Rstar = O(1),\quad\quad 0 < \kappa \ll 1.
\end{equation}
From the general theory of slender bodies in Stokes 
flow \citep{Cox70,Batchelor70,Keller+Rubinow}, we know that
to leading order in an expansion in $1/\ln\kappa$,
\begin{equation}
\Kt \sim {R}^{-1} (\alpha \proj{z} + \beta \proj{\perp}),
\end{equation}
where both $\alpha$ and $\beta$ are $O(1/\ln\kappa)$.
In terms of components,
\begin{equation}
\label{eq:slender-Kt-components}
\kt_{zz} \sim {\alpha}{R}^{-1},
\,\,\,
\kt_{\rho\rho} \sim \kt_{\phi\phi} \sim {\beta}{R}^{-1}.
\end{equation}
The symbol `$\sim$' is used in the asymptotic analysis sense; in the
present case it means that the difference between a left-hand and
right-hand expression vanishes faster than $1/\ln\kappa$ as $\kappa\to 0$.

Inserting the expressions (\ref{eq:slender-Kt-components}) 
into (\ref{eq:U-final}), and replacing longitudinal arc 
length $\ell$ by axial coordinate $z$, which is legitimate up to a
correction of relative order $\kappa^2$, yields
 \begin{equation}
\label{eq:slender-body-V}
{\bm U}_{\text{p}} \sim 
\frac{\mu}{{\mathcal D}}
\int \left(
\kappa \frac{d\Rstar}{dz}J_0(z) \evec{z} 
-\frac{1}{2}{\bm J}_\perp(z) 
\right)
 \, \frac{dz}{2}. 
  \end{equation}
These are the leading-order contributions to each component, in an expansion in
$|\ln \kappa|^{-1}$. That the axial component is $O(\kappa)$, while the
transverse is $O(1)$ comes from (\ref{eq:U-final}), not from the asymptotic
expression for the surface traction. 
A special case of (\ref{eq:slender-body-V}), that of an axisymmetric flux distribution 
(${\bm J}_\perp=0$), has been derived previously \citep{Yariv-11,Schnitzer+Yariv-15}.
Note that, since only the $z$-component depends on $\kappa$ at fixed flux density,
if ${\bm J}_\perp$ is nonzero, the velocity will be nearly transverse
for small enough $\kappa$.

We now consider rotation about an axis in the transverse plane and containing the origin. 
In the slender body limit, the force generated on a short segment of the body at 
$z$ is equivalent to that for a pure translation with velocity ${\bm \Omega} \times {\bm r}$
because the velocity is nearly uniform when $z$ varies of order $\kappa$.
Such reasoning clearly does not work for rotation about the $z$-axis, but we know from the
discussion in Section \ref{sec:symmetry} that we need not consider such rotation.
Thus, 
\begin{equation}
\Kr{\bm\Omega} 
\sim \Kt ({\bm\Omega} \times {\bm r}) 
= (-\Kt\times {\bm r}) {\bm\Omega},
\nonumber
\end{equation}
which gives
\begin{equation}
\Kr  \sim -\Kt\times {\bm r}
\, \sim \, {\beta} \frac{z}{R}(\evec{\rho}\evec{\phi} - \evec{\phi}\evec{\rho}). 
\nonumber
\end{equation}
The integral of this last expression over the surface is zero, verifying that
asymptotically, the center of resistance is located at the coordinate origin.
Insofar as the force on a length $dz$ of the body is independent of $\Rstar$,
and therefore the torque on said segment depends only on $z$, this was actually
fairly obvious. But, to find ${\bm\Omega}$ we do need the leading-order components
\begin{equation}
  \kr_{\rho\phi} = -\kr_{\phi\rho} \sim \beta{z}/{R}.
\nonumber
\end{equation}
Applying (\ref{eq:Omega-final}) now gives
  \begin{equation}
\label{eq:slender-body-Omega}
{\bm\Omega}_{\text{p}} \sim 
-\frac{3}{4} 
\frac{\mu}{{\mathcal D}}
\int \evec{z}\times 
z {\bm J}_\perp(z) \, {dz}.
  \end{equation}

\section{Concluding remarks}

The slender body results (\ref{eq:slender-body-V}, \ref{eq:slender-body-Omega})
promise to be good only to within corrections of relative order $1/\ln\kappa$. 
\citep{Schnitzer+Yariv-15} showed that for the particular slender body family 
comprised of highly eccentric spheroids, the corrections to 
the axial component of ${\bm U}_p$ are actually algebraic.
We now see that the correction is even of relative order $\kappa^2$. 
For, the only differences between the spheroid formulas and the 
slender body formulas are the factor $dz/d\ell = 1 + O(\kappa^2)$ inside the 
integrals (\ref{eq:spheroid-V}, \ref{eq:spheroid-Omega}) and the prefactor 
$(1-\kappa^2)/(1+\kappa^2) \sim 1 - 2\kappa^2$.
An interesting aspect of the slender body velocity (\ref{eq:slender-body-V})
for the case that ${\bm J}_\perp\equiv 0$ is the factor of $d\Rstar/dz$.
On its face, this says that flux on the {\em sides} of a cylinder is 
completely ineffective, and only the ends contribute to motion.
This harsh verdict may be mitigated by deviation from the 
slender body limit or, more likely, by significant thickness of the
interfacial layer. 
More interestingly, it says that a shape which is pinched near the 
middle of its length, with flux of opposite signs on the two ends,
but only on the parts where $R$ increases moving away from the center
(inert endcaps) will go backward with respect to expectations based 
on experience with fully convex motors. 
It may be that the phenomenon is not intrinsically linked to the 
slender body limit. In that case, the most experimentally accessible
geometry may be a pair of fused Janus spheres, with the active hemispheres 
facing each other.

\begin{acknowledgments} 
This work was supported by the NSF under grant DMR-1420620
through the Penn State Center for Nanoscale Science.
\end{acknowledgments} 


\end{document}